\documentclass[aps,pra,showpacs, twocolumn, floatfix,amssymb
]{revtex4-2}
\usepackage[toc,page]{appendix}
\usepackage[british]{babel}
\usepackage[utf8]{inputenc}
\usepackage{amsmath,amssymb}
\usepackage{graphicx}
\usepackage{ae}
\usepackage{esdiff}
\usepackage{braket}
\usepackage{bbm}
\usepackage{simplewick}
\usepackage{float}
\usepackage{braket}
\usepackage{simplewick}
\usepackage{color}
\usepackage{framed}
\usepackage[caption=false]{subfig}
\usepackage[pdftex,colorlinks=true,linkcolor=blue,citecolor=blue,urlcolor=black]{hyperref}
\usepackage{hyperref}
\usepackage{mathtools}
\usepackage{enumitem}
\usepackage{ragged2e}
\usepackage{floatflt}
\usepackage{import}
\usepackage{titlesec}
\usepackage{multirow}
\usepackage{etoolbox}
\usepackage{appendix}
\usepackage{dsfont}
\usepackage{comment}
\usepackage{xcolor}
\usepackage{blindtext}
\usepackage[makeroom]{cancel}
\usepackage{mathtools}
\usepackage{lineno}
\usepackage{ulem}

\begin{document}

\newcommand{\abs}[1]{\mathopen|#1\mathclose|}

\newcommand{\red}[1]{{\color{red}{#1}}}
\newcommand{\blue}[1]{{\color{blue}{#1}}}
\def \k{{\mathbf{k}}}
\def \p{{\mathbf{p}}}
\def \q{{\mathbf{q}}}
\def \r {{\mathbf{r}}}
\def \Q {\mathbf{Q}}
\def \R {\mathbf{R}}
\def \x {\mathbf{x}}
\def \y {\mathbf{y}}
\def \a {\mathbf{a}}
\def \b {\mathbf{b}}
\def \d {\mathbf{d}}
\def \c {\mathbf{c}}
\def \bn {0}
\def \intphi { \int \frac{d\phi_k}{2\pi} \frac{d\phi_p}{2\pi}}

\title{Bose-enhanced relaxation of driven atom-molecule condensates} 

\author{Dimitri Pimenov}
\author{Erich J.\ Mueller}
\affiliation{Department of Physics, Cornell University, Ithaca, New York 14853, USA}

\begin{abstract} 
Motivated by recent experiments \cite{chicagoexp} we study the interconversion between ultracold atomic and molecular condensates, quantifying the resulting oscillations and their slow decay.  We find that near equilibrium the dominant damping source is the decay of condensed molecules into non-condensed pairs, with a pair kinetic energy that is resonant with the  frequency of the oscillating atom-molecule interconversions. The decay, however, is non-exponential, as strong population of the resonant pairs leads to   Bose enhancement. Introducing an oscillating magnetic field, which periodically modulates the molecular binding energy, enhances the oscillations at short times. However, the resulting enhancement of the pair-production process results in an accelerated decay which rapidly cuts
off the initial oscillation growth. 
\end{abstract}

\maketitle

\section{Introduction}

At ultracold temperatures atoms and molecules can coherently interconvert \cite{RevModPhys.78.1311}. For bosonic atoms that can be manipulated via magnetically tunable Feshbach resonances, this enables the experimentally controlled creation of atomic and molecular Bose-Einstein condensates (BECs) and their coherent superposition, a cornerstone of ``ultracold superchemistry" \cite{PhysRevLett.84.5029, 10.1063/1.4964096}. Recent experiments involving magnetic field quenches near a very narrow Cs resonance \cite{zhangfirst, chicagoexp,wang} have been able to explore this phenomenon in a detail that was not possible with earlier studies relying upon Ramsey-type interferometry near broader Feshbach resonances \cite{Donleyexp}.
 These new experiments observed slowly damped oscillations of the atom-molecule BEC fractions, directly visualizing the macroscopic coherence.  They further demonstrated that by periodically modulating the magnetic field these oscillations could be enhanced.  Here we model these phenomena.  We produce a microscopic theory of the damping and present a framework for understanding the driven system.

The coherent atom-molecule interconversion process is somewhat analogous to the quantum mechanical motion of particles in a double-well potential.  One well represents unpaired atoms and the other paired molecules.  The system is, however, highly nonlinear. Depending on the detailed form of the interactions, and the chosen initial state, one finds complicated dynamics already at the mean-field level \cite{PhysRevLett.84.5029, PhysRevA.59.R3186, PhysRevLett.86.1915, PhysRevA.73.023609}. 
While a full theoretical description of the time evolution is challenging, a reasonable expectation for the long-time limit  is that the system relaxes to the global minimum of the mean-field energy landscape, which is a fixed point of the corresponding equations of motion.  Indeed, in Ref.\ \cite{chicagoexp} the long-time dynamics were well described by damped oscillations, and the asymptotic populations  were  used to constrain the form of the atom-molecule coupling.  Our primary goal is to model the damping in these experiments.  We caution that the system likely also supports long lived metastable states, which could be reached in these dynamics.  Thus in Sec.~\ref{sec:exp} we advocate for somewhat more controlled experiments which are designed to explore the oscillations close to equilibrium.

Our approach is to write down a variational wavefunction which includes both condensed and non-condensed degrees of freedom, corresponding to a variant of a Bogoliubov approximation \cite{PhysRevLett.86.1915, PhysRevLett.88.090403,  PhysRevLett.89.180403, PhysRevLett.100.093001, PhysRevA.73.043611}.
An important feature of this modeling is that it explicitly tracks the %relevant non-condensed 
degrees of freedom which are responsible for the dissipation.  This approach can be contrasted with treatments that {\it integrate out} the non-condensed modes, producing an effective model \cite{PhysRevA.31.2403, PhysRevLett.79.6, rivas2012open}.  While elegant, eliminating {the non-condensed modes} relies on crucial approximations which break down under the experimental conditions. First, a separation of time-scales is needed: the bath of non-condensed atoms must equilibrate on a time-scale that is short compared to those that govern the condensate dynamics. Second, these approximations require that the properties of the bath are not influenced by the condensate dynamics. These underlying assumptions are commonly known as ``Markovian".

In the large density limit, where the interaction with the condensate particles dominates over the inter-bath interactions, 
 the Markovian assumption fails. {
 A number of  Bogoliubov approximations have been developed  \cite{PhysRevLett.86.1915, PhysRevLett.88.090403,  PhysRevLett.89.180403, PhysRevLett.100.093001, PhysRevA.73.043611} to circumvent this challenge and include the leading order fluctuations. In the resulting differential equations, time-dependent continuum fields are retained explicitly (i.e., the bath is not integrated out).  This greatly  complicates  the study of general  trajectories. }

In this work, we extend these Bogoliubov-type descriptions of continuum effects to model the new experiments \cite{zhangfirst, chicagoexp} which are, for the first time, able to observe real-time atom-molecule oscillations, the damping of these oscillations, and the influence of periodic driving on them. To this end, we  use our variational wavefunction to analyze the universal long-time  coherent oscillations of atomic/molecular BECs  near their equilibrium populations. We find that the oscillations are damped by the coupling to the continuum in a highly non-trivial manner: the oscillations cause a resonant decay of molecules into correlated atom pairs with a kinetic energy that is set by the oscillation frequency. At shorter times this results in exponential damping; at longer times the damping becomes superexponential due to a Bose-enhancement of the resonantly populated continuum mode. This enhancement is particularly virulent when the system is externally driven at its natural oscillation frequency, since the overpopulation of the continuum modes rapidly cuts off any initial oscillation enhancement from the drive, in line with experimental observations \cite{chicagoexp}. 

 In analyzing their experiment, the authors of Ref.\  \cite{chicagoexp} made the case that they were observing a process where three atoms collide, leaving behind a molecule and an atom. While we focus on simpler two-body interconversion processes, we expect our results to qualitatively translate to this situation as well{, as we briefly discuss in App.\ \ref{app:3body}. On general grounds, we expect that three-body processes, while important for high-energy non-equilibrium trajectories, are less crucial for equilibrium properties, since the three-body interaction becomes irrelevant in the low-energy limit in a renormalization group sense. Although we can only briefly touch on it here, we furthermore find that the Bogoliubov description  of three-body processes comes with an interesting technical subtlety: at the energy minimum of the system, an extensive population of atomic pairs with vanishing momenta becomes favorable, a so-called pair condensate. In a simpler  context, Nozier\`es and Saint James \cite{nozieres1982particle} have argued that a pair condensate only appears for unphysical interaction parameters which result in a mechanical instability (collapse). We leave the detailed analysis of the physical viability of pair condensation due to three-body interactions for future work. }

During the preparation of this manuscript a closely related analysis was carried out by Wang et al.\ \cite{wang}. These authors considered a similar variational ansatz and compared their numerics to both the data from \cite{chicagoexp} as well as new experimental data.  Despite a similarity in methods, there is little overlap with our work, and   the analysis in Ref.\  \cite{wang} complements our  own: They concentrate on the experimental ``quench'' protocol, where the system is thrown far from equilibrium, while we are  concerned with the oscillatory dynamics near the fixed point. We give a microscopic explanation of how pair production leads to the decay of coherent atom-molecule oscillations, identifying the relevant modes and how they evolve.  We show how 
Bose enhancement leads to non-exponential damping and find 
signatures in the population of non-condensed particles.
We show that these effects are further enhanced when the system is driven by a modulation of the magnetic field.   Wang et al.\  instead give rich insights into large amplitude atom-molecule oscillations, and how they depend on experimental parameters.

The remainder of this article is structured as follows: In Sec.\ \ref{energeticssec} we analyze the energy landscape  within the coherent-state approximation. In Sec.\ \ref{dynamicssection} we describe the dynamics: after reviewing the possible mean-field trajectories and commenting on finite particle-number deviations from mean field, we model the damped BEC oscillations close to the minimal-energy fixed point. We  argue that the dominant source of damping is the resonant decay of molecules into excited pairs, derive the corresponding damping rate, and analyze the long-time behavior where the resonant pairs become strongly populated. In Sec.\ \ref{drivingsec} we analyze the impact of the experimentally implemented periodical modulation of the molecule level. 
 In Sec.~\ref{sec:exp} we discuss experimental considerations:  Energy and time scales, and how experimental protocols could be modified to explore our results.
A conclusion and outlook are presented in Sec.\ \ref{conclusionsec}. Various technical derivations and numerical visualizations are relegated to the Appendices.

\section{Energetics}
\label{energeticssec}
We consider an ensemble of bosonic atoms and molecules in three spatial dimensions at $T=0$. If we neglect the effects of the optical trap, the system is governed by the model Hamiltonian
\begin{align}
\label{basichamil}
H  = &\sum_k\left(\epsilon_0+\frac{k^2}{4m}\right) \phi_k^\dagger \phi_k
+ \frac{k^2}{2m} \psi_k^\dagger \psi_k  \\ 
+ \frac{\lambda}{\sqrt{V}} 
&\sum_{pq} \phi_{p+q}^\dagger \psi_p\psi_q
+ \psi_p^\dagger\psi_q^\dagger \phi_{p+q}\  , \notag
\end{align} 
where the operator $\phi_k (\psi_k)$ annihilates a molecule (atom) with momentum $k$.   Here $\epsilon_0$ is the molecular  energy  relative to the atomic continuum, which is chosen to be the zero of energy.  Further,    $\lambda$ is an intensive short-range coupling constant, and $V$ the system volume.  In addition to the Yukawa term, one can also include various density-density interaction terms \cite{wang}.  At resonance we expect those contributions to be sub-dominant, but they can be important for various {out-of-resonance} phenomena  and have to be considered for an accurate determination of  phase boundaries \cite{basu}. 

We use units such that $\hbar =  1$,  take the atomic mass to be $m$ and the molecular mass to be $2m$.  We discuss the numerical values of the parameters in Sec.~\ref{sec:exp}.  In the bulk of the paper we are more concerned with trends and scaling behavior than matching experimental observations.

\begin{figure}
\centering
\includegraphics[width=\columnwidth]{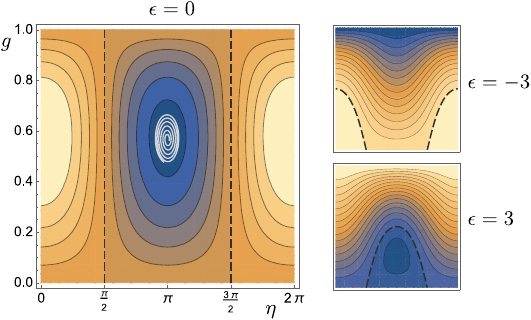}
\caption{{Condensate energy density $E_\text{cond}$ corresponding to Eq.\ \eqref{Energydensity} for three values of the molecular detuning $\epsilon$. Dashed lines correspond to zero-energy contours.  Darker colors represent lower energies. The  spiral in the left figure shows a typical trajectory when the excited atomic states are taken into account, see Sec.\  \ref{dynamicssection}. Energy and timescales used for dimensionalizing are described in Sec.~\ref{sec:exp}.}}
\label{energyfig}
\end{figure}

Let $N = \sum_k \braket{\psi^\dagger_k \psi_k} + 2\braket{\phi^\dagger_k \phi_k}$ be the conserved total number of particles. In the thermodynamic limit $N \rightarrow \infty$, the system is  dominated by the atomic and molecular condensates in the $k=0$ modes. 
 As a first approximation one can neglect
 the population of finite momentum states.  The resulting mean-field physics is captured by a variational coherent state ansatz \cite{10.1063/1.4883893}: 
\begin{align} 
\label{coh1}
\ket{\Psi_\text{cond}} \equiv &\exp\left(- N f^2/2  -N g^2/4\right) \times \\ & \notag \exp\left(z_f\sqrt{N} \psi^\dagger_0 + z_g \sqrt{N/2} \phi_0^\dagger\right) \ket{\text{vac}} \ . 
\end{align} 
Here $z_f = fe^{i\theta}$ and $z_g = ge^{i\chi}$ are complex numbers parametrizing the atomic and molecular condensates, respectively.  They fulfill $f^2 + g^2 = 1$. With the help of the states $\ket{\Psi_\text{cond}}$, we can determine the energy density $E_\text{cond}$ \cite{10.1063/1.4883893}. It is convenient to measure energies in units of  $\sqrt{2n}\lambda$, where $n=N/V$ is the density. For a given renormalized detuning  $\epsilon \equiv  \epsilon_0/(\sqrt{2n} \lambda)$, we can then parametrize $E_\text{cond} \equiv \braket{\Psi_\text{cond}| H |\Psi_\text{cond}}/(\sqrt{2n}\lambda N)$ in terms of the molecular amplitude $g$ and the relative phase $\eta = 2\theta - \chi$. We find 
\begin{align} 
\label{Energydensity}
E_\text{cond} = \frac{\epsilon}{2} g^2 + \cos(\eta) (1-g^2) g \ . 
\end{align} 
An illustration of the energy landscape is shown in Fig.\ \ref{energyfig}.
The energy is minimized  by taking $\eta=\eta^*$ and $g=g^*$ with
\begin{align} 
\label{gstarcond}
 \eta^\star = \pi,  \quad  g^\star  = \frac{\sqrt{\epsilon^2 + 12} -\epsilon}{6} \ . 
 \end{align} Therefore, the ground state is a coherent superposition of atomic and molecular condensates.  The molecular condensate fraction is small but non-vanishing at large positive detunings $\epsilon$.  For $\epsilon \leq-2$, the condensate  is purely molecular.

Now we include the excited states, considering processes
%. To begin with, we neglect excited molecular states and focus on processes 
where a condensed molecule can decay into excited atomic pairs with momenta $k, -k$ (equivalent to ``rogue photodisassociation" discussed 
 by Javanainen and collaborators \cite{PhysRevLett.88.090403, PhysRevA.62.063616}). Such pair creation  is incorporated into our ansatz by extending the coherent state ansatz of Eq.\ \eqref{coh1} to {\cite{nozieres1982particle}}
\begin{align} 
\label{newpsi}
&\ket{\Psi}\equiv  \frac{\exp\left(- N f^2/2  -N g^2/4\right)}{\sqrt{\mathcal{N}}} \times  \\ \notag &\exp\left(z_f\sqrt{N} \psi^\dagger_0 + z_g \sqrt{N/2} \phi_0^\dagger + e^{i\chi}\sum_{k}{\vphantom{\sum}}' z_k \psi_k^\dagger \psi_{-k}^\dagger\right) \ket{\text{vac}},  \\ & 
\mathcal{N} = \prod_{k}{\vphantom{\sum}}' (1-|z_k|^2)^{-1} \ . \notag 
\end{align} 
{Here, the prime 
{in sum and product} indicates that only half of all momenta need to be taken into account to avoid double counting { in the sums and products} \cite{nozieres1982particle}, which can for instance be implemented by imposing $k_x \geq 0$. } In addition to $z_g, z_f$, the generalized coherent states are parametrized by momentum-dependent complex numbers $z_k$, with an extra factor of $e^{i\chi}$ pulled out for convenience.

As  is common to  many variational descriptions, the accuracy of the anstatz \eqref{newpsi} is hard to quantify. It becomes exact in the condensate-only limit and, in the spirit of Bogoliubov theory, it should capture the leading high-density corrections.  The wavefunction  $\ket{\Psi}$ has the properties (see App.\ \ref{cohapp}):  
\begin{align} 
\label{Psiidentities}
\braket{\Psi |\psi_k \psi_{-k} |\Psi} = \frac{z_k e^{i\chi}}{1-|z_k|^2} \ , \quad 
\braket{\Psi |\psi_k^\dagger \psi_{-k} |\Psi}  = \frac{|z_k|^2}{1-|z_k|^2} \ . 
\end{align}

\begin{figure}
\centering
\includegraphics[width=\columnwidth]{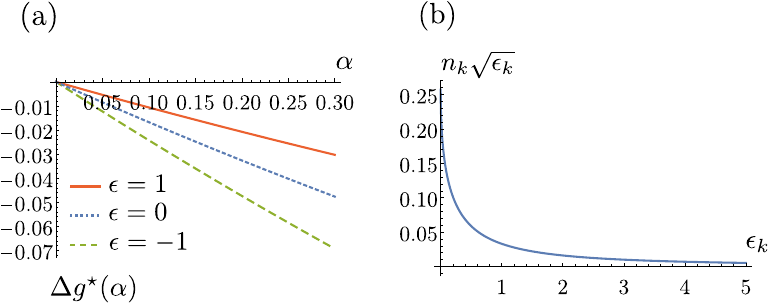}
\caption{(a) Shift  of the equlibrium molecular amplitude due to coupling to the atomic continuum: $\Delta g^\star(\alpha) = g^\star(\alpha) - g^\star(0)$. (b)  Equilibrium atomic mode occupation $n_k$ weighted by the density of states $\sqrt{\epsilon_k}$, Eq.\ \eqref{nkeq}, for $\epsilon = 0$ and $\alpha \rightarrow 0$. }
\label{minplot}
\end{figure}

Taking into account the excited states, particle number conservation implies 
\begin{align} 
\label{PNconservation}
f^2 + g^2 + \frac{1}{N} \sum_k \frac{|z_k|^2}{1- |z_k|^2}  = 1.  
\end{align} 
In a spatially homogeneous system, momentum-dependent quantities can be expressed as functions of the scaled energies 
\begin{align} 
\epsilon_k \equiv \frac{k^2}{2m} \times \frac{1}{\lambda \sqrt{2n}}
\ . 
\end{align} 
Going to the continuum limit, the particle number conservation can then be rewritten as 
\begin{align}
\label{pncons}
&f^2 + g^2 + \alpha \int_k \frac{|z_k|^2}{1- |z_k|^2}  =  1 \ ,   \\ 
&\int_k \equiv \int d\epsilon_k \sqrt{\epsilon_k} , \quad  
%\alpha = \frac{m^{3/2}}{\sqrt{2} \pi^2 \rho} \times \left( \lambda \sqrt{2N}\right)^{3/2}\ ,
\alpha = \frac{m^{3/2}}{\sqrt{2} \pi^2 n} \times \left( \lambda \sqrt{2n}\right)^{3/2}.
\label{alphadef}
\end{align} 
%where $\rho = N/V$ is the particle density. 
 Here
$\alpha$ is the effective dimensionless coupling constant in the problem, which vanishes in the high-density or weak-interaction limit. The energy density 
$E \equiv \braket{\Psi | H |\Psi}/(\sqrt{2}N n^{1/2} \lambda)$
becomes
\begin{align} 
\label{Ewithalpha}
E = &\frac{\epsilon}{2} g^2 +\cos(\eta) g\left(1 - g^2 - \alpha \int_k \frac{|z_k|^2}{1- |z_k|^2} \right)    \\+ &\alpha \int_k \frac{1}{1 - |z_k|^2} \left( \epsilon_k |z_k|^2 + u_k  g\right) , \notag
\end{align} 
where $z_k = u_k + i v_k$. At the energy minimum  the phase difference satisfies $\eta = \pi$, as in the condensate-only case. The minimal-energy choice for the excited state coefficients is 
\begin{align} 
\label{vkmin}
v_{k}^\star &= 0   \\ 
\label{ukmin}
u_{k}^\star(g) &=  \frac{-(g + \epsilon_k) + \sqrt{(g + \epsilon_k)^2 - g^2}}{g}   \\  &%=
\to
\begin{cases}
-1, \ &\epsilon_k \rightarrow 0 \\ 
 - \frac{g}{2\epsilon_k} &\epsilon_k \rightarrow \infty \notag \ . 
\end{cases} 
\end{align} 
Eq.\ \eqref{vkmin} implies that phase-locking between the molecular condensate and the excited atomic wave-functions is energetically favorable. Although {$1/(1-(u_k^*)^2)$} diverges as $k\to0$, the divergence is integrable.  However, when taken at face value, the  ultraviolet behavior of $u_k^\star$ is problematic: inserting Eq.\ \eqref{ukmin} back into Eq.\ \eqref{Ewithalpha}, the second integral on the r.h.s. is divergent, {nominally implying  $E \rightarrow - \infty$}.  This unphysical divergence is an artifact of the short-range coupling, and can be averted by introducing a renormalized detuning $\epsilon_\text{ren}$  \cite{PhysRevLett.86.1915,PhysRevLett.89.180403}:
\begin{align} 
\epsilon_\text{ren}&\equiv \epsilon - \alpha\int_k \frac{1}{2\epsilon_k}  \\ 
\label{Ewithalpharen}
E = &\frac{\epsilon_\text{ren}}{2} g^2 +\cos(\eta) g\left(1 - g^2 - \alpha \int_k \frac{|z_k|^2}{1- |z_k|^2} \right)    \\+ &\alpha \int_k \left[\frac{1}{1 - |z_k|^2}  (\epsilon_k |z_k|^2 + u_k  g)+ \frac{g^2}{4\epsilon_k}\right] \ . \notag 
\end{align} 
In the following, we will drop the subscript in $\epsilon_\text{ren}$ for brevity. Inserting $u_k^\star(g)$ into Eq.\ \eqref{Ewithalpharen}, we can numerically determine $g^\star$ as a function of $\alpha$. The result is shown in Fig.\ \ref{minplot}(a). We see that introducing the excited states leads to an analytic correction $g^\star(\alpha)= g^\star(0) - c_\epsilon \alpha+\mathcal{O}(\alpha^2)$, where $c_\epsilon$ is a detuning-dependent $O(1)$ constant  and the higher order in $\alpha$ terms are negligible on the scale in the figure. In Fig.\ \ref{minplot}(b), we show the corresponding mode occupation weighted by the density of states, 
\begin{align} 
\label{nkeq}
\sqrt{\epsilon_k} n_k = \sqrt{\epsilon_k} (u_k^\star(g^\star))^2/(1-(u_k^\star(g^\star))^2) \ . 
\end{align} 
Note that the total continuum occupation, $N_k = \alpha \int_k n_k$, is weighted by the coupling constant $\alpha$ and therefore vanishes as $\alpha\to 0$.
 The atomic and molecular condensates are robust against adding these modes, and in the high density ($\alpha\to 0$) limit we recover the mean-field results that came from the ansatz in Eq.~(\ref{coh1}).

In principle,  our ansatz, Eq.~\eqref{newpsi}, can be extended to include excited molecular states, 
for example by considering an expression of the form
\begin{align}
\ket{\Phi}
 =  \exp\left(
\sum_k \eta_k \phi_k^\dagger \psi^\dagger_{-k}
+
\sum_{k}{\vphantom{\sum}}'  w_k \phi_k^\dagger \phi^\dagger_{-k}
\right)
|\Psi\rangle.
\end{align}
Here $\eta_k$ and $w_k$ are related to the presence of correlated atom-molecule or molecule-molecule pairs.
In Ref.~\cite{wang}, the authors included $w_k$, but not $\eta_k$, finding that adding those terms had no impact on the dynamics.  This result is sensible, as there are no terms in the Hamiltonian which can produce correlated molecule-molecule pairs without first forming atom-molecule pairs.  Thus if $\eta_k=0$ one should also have $w_k=0$.  

{In App.\ \ref{excitedmolapp}, we extend our ansatz $\Psi$ by including the $\eta_k$-terms; we expect the $w_k$-contributions to be less important at small densities of excited molecules. We find that  extending our ansatz  in this way  leads to an energy minimum that contains a pair condensate \cite{nozieres1982particle}, similar to the case of a  three-body interaction (App.\ \ref{app:3body}). It remains to be seen whether this pair condensate survives the inclusion of stabilizing molecule-molecule interactions. }

Regardless, we do not expect that the molecular continuum has a major impact  on experimental timescales. The leading scattering process $\phi^\dagger_k \psi_k \psi_0$ which produces excited molecular states involves excited atomic states, whose occupation is considerably suppressed relative to the condensate modes {in the controlled high-density limit $\alpha \ll  1$ we are interested in}. Therefore, following previous literature, we neglect the molecular continuum in this article; its correct treatment will be the subject of future work \cite{threebody}. {In the high-density limit, we may also neglect further coupling between the excited modes, which could in principle be captured by considering variational ansaetze with more than two non-condensed operators in the exponent. 
 At sufficiently long times the momentum distributions should thermally equilibrate, a process which is often modeled using a quantum Boltzmann equation (for example see \cite{vaibhav}).} 
 
\section{Dynamics}
\label{dynamicssection}

We proceed with  calculating the system dynamics,  applying the time-dependent variational principle \cite{10.1063/1.431911}. It is based upon the fact that the many-body wave function which  solves the Schroedinger equation is a stationary point of the action
\begin{equation}
S=\int dt\braket{\Psi(t)| i\partial_t-H|\Psi(t)} \ . 
\end{equation}
We use the coherent-state ansatz, Eq.~\eqref{newpsi}, for $\ket{\Psi(t)}$, with time-dependent coefficients $z_f(t), z_g(t), z_k(t)$. The stationary-point condition,  $\delta S/\delta \bar{z}_i(t)=0$,  where $\bar z$ is the complex conjugate of $z$, leads to the following set of equations (see App.\ \ref{diffeqapp}): 
\begin{align}
\label{fulltwo}
\dot{g} &= \sin(\eta) f^2 + \alpha \int_k \frac{v_k}{1-|z_k|^2} ,  \\ \notag 
\dot{f} &= - fg 
\sin(\eta),  \\ \notag
\dot{\eta} &= \epsilon + \cos(\eta) \left( \frac{f^2}{g} -2g\right) + \alpha \int_k \left( \frac{u_k}{g(1-|z_k|^2)} + \frac{1}{2\epsilon_k} \right)  ,  \\ \notag
i\dot{z}_k &= 2 z_k(\epsilon_k - \mu)  + g(1 + z_k^2) \ . 
%\dot{u}_k &= 2(\epsilon_k - \mu) v_k + 2 g u_k v_k  \\ \notag 
%\dot{v}_k &= -2(\epsilon_k - \mu) u_k -g - g(u_k^2 - v_k^2)\  .
\end{align}
 The time-dependent chemical potential $\mu$ is
\begin{align}
\mu &= \frac{\epsilon}{2} + \frac{f^2}{2g}\cos(\eta) + \frac{\alpha}{2} \int_k \left( \frac{u_k}{g(1 - |z_k|^2)} + \frac{1}{2\epsilon_k}\right), 
\end{align}
which determines the population of continuum states and arises from the time-dependence of the molecular condensate phase, $\mu = - \dot{\chi}/{2}$. Eq.\ \eqref{fulltwo} is in agreement with the results of Ref.\ \cite{wang}. Related differential equations have also previously been reported e.g. in Refs.\ \cite{PhysRevLett.88.090403,PhysRevLett.89.180403}, based on  approximations of Heisenberg equations of motion where atomic operators are treated as c-numbers. In those prior treatments, the denominators $\sim 1/(1-|z_k|^2)$ are absent, which  are crucial for  correctly describing the long-time evolution. 

In the following we systematically analyze the solutions to Eq.\ \eqref{fulltwo}: 
In Sec.~\ref{mfd} we discuss the mean-field dynamics, and quantum fluctuations in the single-mode limit. 
In Sec.~\ref{aco} we make general remarks about the influence of the noncondensed modes and in Sec.~\ref{linac} we linearize about the fixed point, producing our central results.

\subsection{Mean-Field Dynamics and single-mode fluctuations}\label{mfd}

\begin{figure}
\centering
\includegraphics[width=\columnwidth]{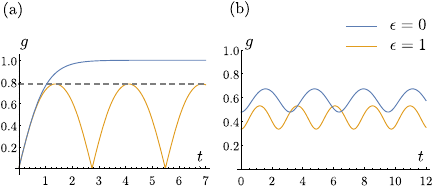}
\caption{Time evolution of molecule population for two different molecular energies $\epsilon$ without coupling to the atomic continuum, $\alpha = 0$. (a) Boundary condition $g(0) = 0$. The dashed horizontal line marks $g_\text{max}$ (main text). (b) $g(0) \lesssim g^\star, \eta \simeq \eta^\star$. }
\label{solitonfig}
\end{figure}

 Before exploring the role of the non-condensed modes, it is important to understand the mean-field dynamics which occur when $\alpha=0$. Absent  dissipation, physical trajectories  follow the equal-energy contours depicted in Fig.\ \ref{energyfig}. Two special cases are of interest. First, one can consider a scenario where at $t=0$ the system consists of atoms only ($f =1, g= 0^+$), which can be accomplished in experiment by quenching the molecular level position from large to small detuning \cite{chicagoexp}. Per Eq.\ \eqref{Energydensity}, the time-evolution proceeds along the zero-energy contour [dashed lines in Fig.\ \ref{energyfig}(a)]. For vanishing molecular level detuning, $\epsilon = 0$, the resulting trajectory  monotonically grows,  asymptoting to $g =1$ as $t \rightarrow \infty$, see Fig.\ \ref{solitonfig}(a) \footnote{Notably, the point $g =1$ is an unstable fixed point of the set of equations \eqref{fulltwo} when $\alpha = 0, \epsilon < 2$. \cite{PhysRevA.62.063616, PhysRevA.64.063611}}. For $\epsilon \neq 0$, $g(t)$ only reaches a value of $g_\text{max} = \sqrt{1+ (\epsilon/4)^2} - \epsilon/4$, and oscillates with a period which scales as $\log(1/\epsilon)$ at small $\epsilon$. This soliton-like trajectory does not depend on the initial condition for the relative phase variable $\eta$ (see App.\ \ref{solitonapp} for details).

 That behavior can be contrasted to the small oscillations which are found when the trajectory begins
 close to the energy minimum 
 at 
$g =  g^\star + \delta g, \eta = \eta^\star + \delta \eta $ [Eq.\ \eqref{gstarcond}]. Linearization leads to the simple oscillator equation 
\begin{align} 
\ddot{\delta g} + \omega_g^2 \delta g = 0 , \quad \omega_g = \frac{\sqrt{1 + 2(g^\star)^2 - 3 (g^\star)^4}}{g^\star} \ , 
\end{align} 
and $g$ oscillates around $g^\star$ with frequency $\omega_g$ \cite{chicagoexp}, see Fig.\ \ref{solitonfig}(b). Note that the oscillation becomes slow, $\omega_g \rightarrow 0$ when the condensate becomes purely molecular at $\epsilon \rightarrow -2$.

The trajectories depicted in Fig.\ \ref{solitonfig} are obtained in the mean-field limit.  That analysis neglects coupling to the continuum, and ignores the quantum fluctuations of the $k=0$ modes. What is their fate beyond these idealized conditions?  While the bulk of our efforts are aimed at understanding the role of the continuum, we briefly comment on the role of $k=0$ quantum fluctuations. These quantum fluctuations lead to decoherence and deviations from the mean-field trajectories at long times $t> t_{\text{MF}}$ \cite{PhysRevLett.86.568,PhysRevA.64.063611, Richter_2015}. When $g(0) = 0$, and $\epsilon \ll 1$, 
 one finds $t_{\text{MF}}\sim \ln N$, and for longer times one observes substantial differences from the soliton-like mean-field solutions
 \cite{PhysRevLett.86.568,PhysRevA.64.063611}. The oscillatory solutions  with $g \simeq g^\star$ are more robust. The decoherence induces some damping;  however, for large systems this effect is very weak since  $t_{\text{MF}} \sim N$ in this case. A comparison of mean-field dynamics and numerically exact results for finite $N$ is presented in App.\ \ref{beyondmfapp}, demonstrating this behavior.  For experimentally relevant system sizes we can ignore the $k=0$ quantum fluctuations.

\subsection{Overview of the atomic continuum}\label{aco}

 The atomic continuum plays two dissipative roles:   over time it  both reduces the number of condensed atoms and pulls energy from the condensate.
 Unlike the decoherence from $k=0$ fluctuations, this damping is relevant even for relatively large $N$, and its magnitude is controlled by the coupling constant $\alpha\propto N^{-1/4}$. The dissipative influence of the non-condensed modes is generic: one encounters similar dissipation whenever a Hamiltonian system is coupled to a bath,  and the damping rate can  formally be obtained by integrating out the bath, as customary in the study of open quantum systems. Commonly, this is done under a Markovian approximation, where the bath equilibrates on the shortest time scale, retaining no memory of interaction with the condensate. 
 Such Markovian treatments of the atom-molecule oscillations can be found in Ref.~\cite{PhysRevA.85.013618} (and references within).
In particular, the authors of Ref.\ \cite{PhysRevA.85.013618} show that the oscillatory (elliptic) fixed point becomes attractive, such that the oscillations die out \cite{PhysRevA.85.013618, Trimborn_2008}. 

 That analysis holds in the low-density limit where the equilibrating inter-bath interactions are larger than the BEC-bath ones. In the high-density limit under study in this work, the Markov approximation breaks down. The compact form of the the differential system \eqref{fulltwo} allows us to retain the continuum degrees of freedom explicitly, and track the precise way the decay occurs in our model. 

{\subsection{Dynamics of soliton-like solutions coupled to the continuum}}

{In Fig. \ref{expdyn}, we present the soliton-like trajectories, with boundary condition $g(0) = 0, f(0) = 1$ as appropriate for quench-type experiments \cite{chicagoexp, wang}. We choose a value of the continuum coupling $\alpha = 0.27$ which reflects current experimental conditions, see Sec.\ \ref{sec:exp}. Our numerics are in good agreement with concurrent results presented in Ref.\ \cite{wang}, where a similar model is employed and detailed comparison to experimental data is presented, highlighting strengths and weaknesses of the model. Generically,  we find that the molecular population relaxes to a stationary value which is  smaller than the equilibrium fixed point value $g^\star$. This can be attributed to a stronger population of the continuum modes during the initial time evolution. We observe a few damped oscillations around the long-time asymptote. The damping is weakest for a the negative molecular detuning $\epsilon = -1$, which is due to the fact that the  molecular state lies outside of the atomic continuum.  Under those conditions the decay process is a {\it many-body} effect, where the energy required to break up a molecule comes from the condensate as a whole.  Understanding this slow damping is the central goal of our  study of the linearized trajectories in Sec.~\ref{linac}.

\begin{figure}
\centering
\includegraphics[width=\columnwidth]{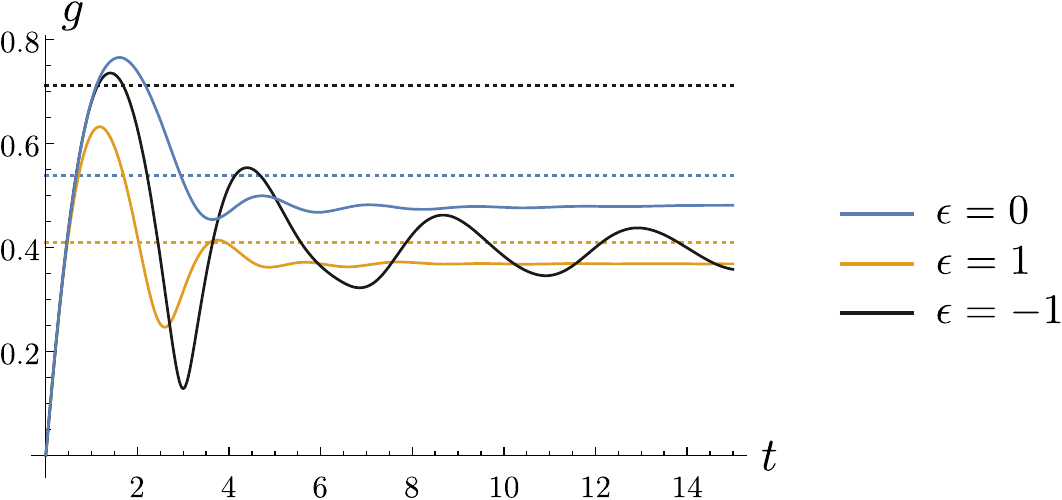}
\caption{{Time evolution of molecule population for boundary condition $g(0) = 0$, continuum coupling $\alpha = 0.27$ and three different molecular energies $\epsilon$. The dashed lines mark the equilibrium fixed point values $g^\star$. }}
\label{expdyn}
\end{figure}

The soliton-like solutions, which have also been studied in several previous works \cite{PhysRevLett.86.1915, PhysRevLett.88.090403,  PhysRevLett.89.180403, PhysRevLett.100.093001, PhysRevA.73.043611, wang}, are high-energy trajectories: as such, they are strongly affected by additional decay sources, e.g.\ from atom-atom collisions, trap effects etc., which are not captured by our model. Therefore, we cannot expect the numerics in Fig.\ \ref{expdyn} to model the experiments quantitatively. For the remainder of this paper, we focus on the oscillatory dynamics near the equilibrium fixed point, which does not suffer from such a theoretical limitation and can be addressed by a modified experimental protocol, see Sec.\ \ref{sec:protocol}. }

\subsection{Dynamics near the fixed point}\label{linac}

We now derive our central results, corresponding to the dynamics near the fixed point.
We start by linearizing Eq.\ \eqref{fulltwo} around its fixed point value, $\x \equiv (g, f, \eta, u_k, v_k) \simeq \x^\star + \delta \x$. {Here, $f^\star$ can be determined  by using the particle number conservation, Eq.\ \eqref{pncons}}.  Using the fact that $v_k^\star = 0, \eta^\star = \pi$, we obtain the following differential equation for $\delta g$: 
\begin{align} 
\label{cleanddotg}
&\ddot{\delta g} +  \omega_g^2 \delta g = \\ & \notag  \alpha \int_k \left(\frac{ \dot{\delta v_k}}{1-(u_k^\star)^2} -  \delta u_k  \frac{ (f^\star)^2 (1+ u_k^\star)^2}{g^\star\left(1-(u_k^\star)^2\right)^2} \right) \ , 
\end{align}
where we have neglected an $O(\alpha)$ correction to the oscillator frequency $\omega_g$. As we will now show, to the leading order in $\alpha$  the term on the r.h.s.\  of Eq.\ \eqref{cleanddotg} is proportional to $\dot{\delta g}$ and therefore serves as a damping source. To prove this, we consider the equations for $\delta v_k, \delta u_k$: 
\begin{align} 
\label{deltaveq}
&\ddot{\delta u_k} + \Omega_k^2 \delta u_k = - \delta g \Omega_k F_k \\ \notag
&\ddot{\delta {v_k}} + \Omega_k^2 \delta v_k = -  \dot {\delta g} F_k , 
\end{align} 
In Eq.\ \eqref{deltaveq}, the we used the labels 
\begin{align} 
\label{OmegakFk}
\Omega_k &= 2(\epsilon_k - \mu^\star + g^\star u_k^\star ), \quad \mu^\star = -g \\  \notag
F_k &= 1 + (u_k^\star)^2 - u_k^\star \frac{(g^\star)^2 + 1}{(g^\star)^2} \ . 
\end{align} 
Eqs.\ \eqref{cleanddotg},\eqref{deltaveq} form a closed system.
 We can estimate the effective damping of $\delta g$ to leading order in $\alpha$ 
 by approximating
all quantities in Eq.\ \eqref{deltaveq} by their values at $\alpha = 0$.  
 Below we formally develop this perturbative expansion.  In this leading order treatment, Eq.\ \eqref{deltaveq} describes  a set of undamped oscillators
subject to a sinusoidal driving force. The drive is resonant when $\Omega_k = \omega_g$, signaling enhanced decay of the condensed molecules into atomic pairs at momenta obeying this resonance condition. At vanishing level detuning $\epsilon$, the resonance occurs at $\epsilon_k\simeq 0.5774$. Importantly, this resonance arises from the oscillatory \textit{dynamics} of $g$: For a static $g = g^\star$, the effective molecule level is below the atomic continuum, as signaled by the negative value of the chemical potential $\mu^\star = - g^\star$, and the decay is switched off. For the oscillating solution, the residual energy stored in the oscillation enables decay into pairs with kinetic energy equal to the oscillation frequency. 

\begin{figure}
\centering
\includegraphics[width=\columnwidth]{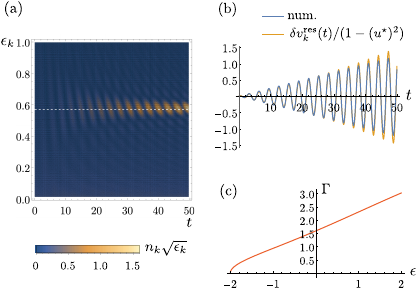}
\caption{Numerical demonstration of the resonant production of atomic pairs during small oscillations about the fixed point. (a) Time-evolution of weighted atomic density $n_k(t) \sqrt{\epsilon_k}$, obtained by discretizing Eq.\ \eqref{fulltwo} on a sufficiently dense momentum grid.  Here we used $\epsilon=0, \alpha=0.01$, and initial conditions with the molecular fraction shifted slightly from its equilibrium value.  {The dashed horizontal line indicates the resonance energy.} (b)  Linear growth of resonant modes illustrated by plotting  $\delta v_k/(1-(u_k^*)^2)$ vs.\ time, for $k$ satisfying the resonant condition (dashed white line in panel (a)).  The full numerical solution (red) is contrasted with the approximation (blue) from Eq.~(\ref{resv}).
(c) Scaled damping rate $\Gamma$ as function of $\epsilon$, see Eq.\ \eqref{Gammaintro}. }
\label{resonancefig}
\end{figure}

In Fig.\ \ref{resonancefig}(a), we plot the time-dependent atomic occupation $n_k = |z_k|^2/(1-|z_k|^2)$ weighted by the density of states $\sqrt{\epsilon_k}$, as obtained from full numerical solution of Eq.\ \eqref{fulltwo} with boundary condition $\x = \x^\star + \delta \x$, $|\delta \x| \ll 1$. The long-time enhancement of the resonant mode is clearly visible.

To obtain  our analytical description of the resonant decay and associated damping,  we parametrize $\delta g(t)\big|_{\alpha = 0} = A\cos(\omega_g t) + B \sin(\omega_g t)$. Close to the resonance one can ignore counter-rotating terms in Eq.~(\ref{deltaveq}) and the solution for $\delta v_k$ can be expressed as 
\begin{align}
\label{resv}
\delta v_k^\text{res}(t) = \  &\frac{B F_k}{2(\omega_g - \Omega_k)} \left( \cos( \omega_g t) - c_1 \cos(\Omega_k t) \right)  \\ \notag -  &\frac{A F_k}{2(\omega_g - \Omega_k)} \left( \sin( \omega_g t) - c_2 \sin( \Omega_k t) \right)  . 
\end{align} 
The coefficients $c_1, c_2$ are determined by boundary conditions,  which generically depend upon the entire history of the system.  Nonetheless we can impose some constraints by assuming  that the non-condensed modes are not macroscopically occupied,  $\delta u_k(0), \delta v_k(0) \ll 1$.  For the near-resonant modes, we then require $c_1 = c_2 = 1$. Similar arguments hold for $u_k$.  The resonance condition in Eq.\ \eqref{resv}  shows that the oscillation frequency $\omega_g$ sets the kinetic energy of excited atomic pairs  which are predominantly generated during the dynamics.  Furthermore, in Eq.~(\ref{resv}) the excitation rate is proportional to $A$ and $B$, which encode the deviation of $g$ from its equilibrium value $g^*$. A numerical check is presented in Fig.\ \ref{resonancefig}(b), where we show $\delta v_k^\text{res}(t)$ weighted by $1/(1-(u^\star)^2)$. This corresponds to the integrand on the right hand side of Eq.\ \eqref{cleanddotg}. 
We see excellent agreement with this perturbative treatment up to times of order $t \simeq 1/|\delta \x|$: at longer times, $\delta v_k^\text{res}, \delta u_k^\text{res}$ become $O(1)$, and the linearization used to derive Eq.\ \eqref{resv} is no longer justified. 

With Eq.\ \eqref{resv} at hand, we can evaluate the right hand side of Eq.\ \eqref{cleanddotg} in the intermediate-time limit, taking into account the near-resonant contributions only (App.\ \ref{rateapp}); the non-resonant terms are rapidly oscillating functions of $\epsilon_k$, and their contributions to the $k$-integral are small. The time-scale on which the resonant contributions start to dominate scales as $1/\omega_g$. The evaluation of the damping rate within the linearized theory is therefore controlled in the large time window $1/\omega_g < t < 1/|\delta \x|$. For these times, we obtain 
\begin{align} 
\label{Gammaintro}
 \alpha \int_k \left(\frac{ \dot{\delta v_k}}{1-(u_k^\star)^2} -  \delta u_k  \frac{ (f^\star)^2 (1+ u_k^\star)^2}{g^\star\left(1-(u_k^\star)^2\right)^2} \right) \  \simeq -  \alpha \Gamma \dot{ \delta g}\ , 
\end{align}
with solution 
\begin{align} 
\label{gsol}
g_\text{analyt.}(t) &=  g^\star + \delta g(t)  \\ &= g^\star +  \left( A \cos(\omega_g t) + B\sin(\omega_g t) \right) \exp(-\frac{\alpha \Gamma}{2} t) \ . \notag
\end{align} 
Here $\Gamma$ is a function of $\epsilon$ which has a complicated analytical form; it is plotted in Fig.\ \ref{resonancefig}(c). $\Gamma$ vanishes as $\epsilon \to -2$, while it grows  roughly linearly at larger $\epsilon$. 

\begin{figure}
\centering
\includegraphics[width=\columnwidth]{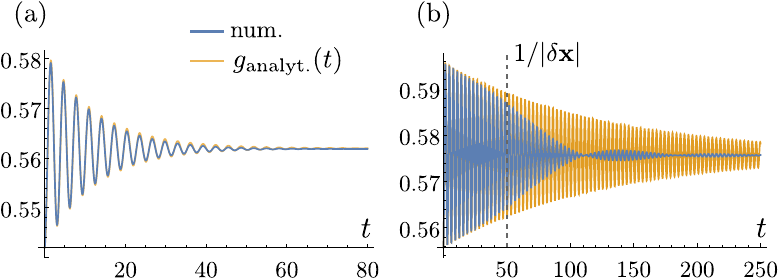}
\caption{Molecular amplitude $g(t)$ at $\epsilon = 0$ for $\alpha = 0.1$ (Fig.~(a)) and a smaller value $\alpha = 0.01$ (Fig.~(b)). In the energy landscape of Fig.\ \ref{energyfig}  the damped oscillation corresponds to a spiral towards the fixed point. The long-time deviation from the ordinary exponential decay, {which sets in at $t \gtrsim 1/|\delta\x|$,}  is clearly observed in Fig.~(b).}
\label{gdecayfig}
\end{figure}

In Fig. \ref{gdecayfig}, we compare the damped oscillation $g_\text{analyt.}$ to the numerically obtained one for a small initial  displacement $|\delta \x| = \delta g(0) = 0.02$. In Fig.\  \ref{gdecayfig}(a), a sizeable overall interaction $\alpha = 0.1$ is used; in this case, the exponentially damped solution Eq.\ \eqref{gsol} agrees with the numerics up to arbitrary times, since the oscillations are already damped out completely for $t > 1/|\delta \x|\sim 50$. 
In Fig.\ \ref{gdecayfig}(b), we consider a smaller value $\alpha = 0.01$.  While we still find excellent agreement for early times, the numerical solution starts to deviate strongly from the approximation \eqref{gsol} at long times  $t\gtrsim 1/|\delta \x|\sim50$, {when the linearization used to derive Eq.\  \eqref{gsol} breaks down. At these longer times, 
much stronger non-exponential damping is observed}. 
This enhanced ``Non-Markovian" damping can be attributed to the strong population of the resonant mode:  a ``Bose enhancement" effect encoded in the $1/(1-|z_k|^2)$ factor in Eq.\ \eqref{fulltwo}. These deviations from the linearized behavior become more important when the oscillations have larger amplitude (i.e., $|{\bf \delta x}|$ is larger) or when the linearized decay rate, $\alpha \Gamma$, is smaller. 
{At the longest times,  small beats are observed in Fig.\ \ref{gdecayfig}(b), which 
 we attribute to
coherent interconversion between condensed molecules and the strongly populated continuum atomic pairs at momenta close to resonance.  }

As described in Sec.\ \ref{sec:exp}, in the current experiments $\alpha\sim 0.27$.  Thus the breakdown of the linearized theory will only occur for large deviations from equilibrium.

Beyond the damping, the strong population of the resonant momentum pair should be observable directly by switching off the trap and imaging the resulting momentum distribution of atoms in a  time-of-flight measurement. This has been achieved in other setups where correlated $\k, -\k$ atom pairs are created, for instance Refs.\ \cite{PhysRevLett.99.150405, BoseFire}.

%\begin{align} 
%\Psi_{\text{cond}}^\text{gen} = \sum_m c_m \ket{N_\phi = m, N_\psi = N - 2m} 
%\end{align} 

\section{Driven systems}
\label{drivingsec}

\begin{figure}
\centering
\includegraphics[width=\columnwidth]{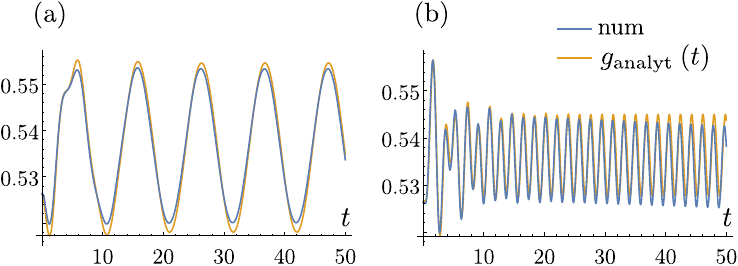}
\caption{ Molecular amplitude $g(t)$ subject to off-resonant drive, $\epsilon(t)=\epsilon+\delta\epsilon \cos(\omega_\epsilon t)$.  Here $\epsilon = 0$,  $\alpha = 0.27$ as in the experiment, see Sec.\ \ref{sec:exp}, and $\delta \epsilon = 0.1$. (a) Driving frequency $\omega_\epsilon = 0.3 \omega_g$. (b) $\omega_\epsilon = 1.7 \omega_g$. The analytical approximation is obtained from solving Eq.\ \eqref{drivenosc}.}
\label{drivenFignonres}
\end{figure}

 The molecule binding energy $\epsilon$ depends on the magnetic field $B$.  Thus  periodically modulating the magnetic field leads to a time dependent $\epsilon$ which can drive the atom-molecule association/disassociation \cite{RevModPhys.78.1311,zhang2022coherent,chicagoexp,PhysRevLett.95.190404,  PhysRevA.81.013602}. 
In  our modeling this drive can be trivially implemented by making $\epsilon$ time-dependent. For simplicity, we focus on the simple sinusoidal driving function, $\epsilon \rightarrow \epsilon + \delta \epsilon \cos(\omega_\epsilon t)$. For the soliton-like trajectories of Fig.\ \ref{solitonfig}(a), periodic driving can enhance the time-averaged molecule population when the driving frequency matches $\epsilon$ \cite{PhysRevLett.95.190404, PhysRevA.81.013602}. Here, we study the impact of the drive on the coherent oscillations close the equilibrium fixed point. For small driving amplitudes $\delta \epsilon$, the linearized equation, Eq.\ \eqref{Gammaintro}, becomes 
\begin{align}
\label{drivenosc}
\ddot{\delta g} + \omega_g^2 \delta g + \alpha \Gamma \dot{\delta g} = - \delta\epsilon \cos(\omega_\epsilon t) (f^\star)^2
\end{align} 
This linearized equation applies as long as $\delta g$ remains small.  This requires that $\delta\epsilon$ is small, and that $\omega_e$ is sufficiently far from the resonant frequency $\omega_g$.  Figure~\ref{drivenFignonres} illustrates this off-resonant regime for an experimentally relevant value of the coupling $\alpha=0.27$:  After an initial transient period, $g(t)$  oscillates with the drive frequency. This figure shows clear agreement between the full numerical calculation and Eq.~(\ref{drivenosc}). The only trend not captured by Eq.~(\ref{drivenosc}) is a small additional decay of $g(t)$ that is observed for the high-frequency drive [Fig.\ \ref{drivenFignonres}(b)] at the longest times; it can be attributed to the population of atomic pairs with a kinetic energy that is commensurate with the driving frequency $\omega_\epsilon$, and is suppressed for smaller values of $\alpha$.

\begin{figure}
\centering
\includegraphics[width=\columnwidth]{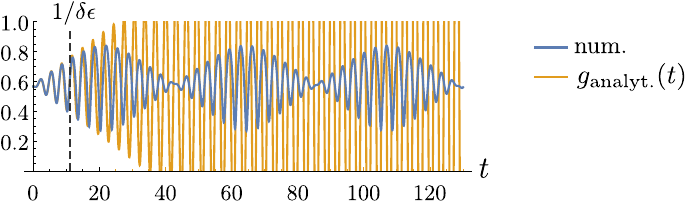}
\caption{Molecular amplitude $g(t)$ subject to resonant drive, $\epsilon(t)=\epsilon+\delta\epsilon \cos(\omega_\epsilon t)$, with $\omega_g \simeq \omega_\epsilon$ without coupling to the continuum, $\alpha = 0$.  Here $\epsilon = 0$, $\delta \epsilon = 0.1$. The analytical approximation $g_\text{analyt.}$, which is valid for short times { $t \lesssim  1/\delta \epsilon$ only}, is  calculated from Eq.\ \eqref{drivenosc}. }
\label{beatfig}
\end{figure}

\begin{figure*}
\centering
\includegraphics[width=\textwidth]{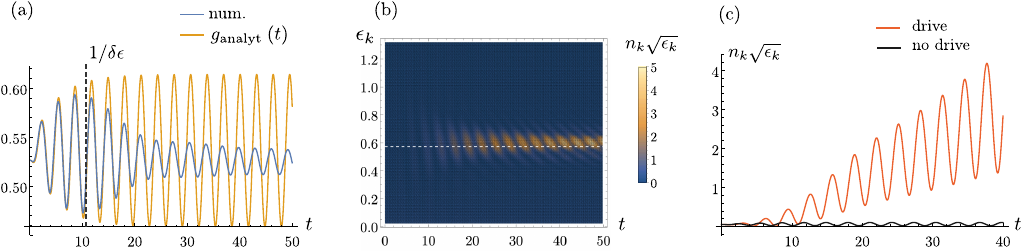}
\caption{(a) Molecular amplitude $g(t)$ subject to resonant drive, $\omega_g \simeq \omega_\epsilon$ with coupling to the continuum. Here $\epsilon = 0$, $\delta \epsilon = 0.1, \alpha = 0.27$. {At times $t > 1/\delta \epsilon$, strong population of the resonant continuum mode sets in, leading to a breakdown of the linearized analytical solution derived from Eq.\ \eqref{drivenosc}}.   (b) Time evolution of excited state population. {The horizontal dashed line marks the resonance energy obtained for $\alpha = 0$.} (c) Time evolution of resonant excited state mode with $\Omega_k \simeq \omega_g$ [see Eq.\ \eqref{deltaveq}], compared to the situation without external drive.}
\label{drivenFigres}
\end{figure*}

For a resonant drive, $\omega_g = \omega_\epsilon$, the simple linearization in Eq.\ \eqref{drivenosc} predicts a linear increase of the oscillation amplitude $\delta g(t)$ as a function of time. However,  once $\delta g$ becomes sufficiently large, the linearized theory breaks down.  Figures~\ref{beatfig} and \ref{drivenFigres} illustrate this breakdown for different parameter values.

In the  limit $\alpha = 0$, when the condensate fractions are decoupled from the continuum, $g(t)$ displays beats when resonantly driven, see Fig.\ \ref{beatfig}. This behavior originates from the fact that the effective potential which determines the $g$-trajectory is  anharmonic. As predicted by the linearized theory, $\delta g$ initially grows linearly in time.  As $\delta g$ grows, however, the natural oscillation frequency shifts out of resonance with the drive.  The resulting phase lag causes the drive to remove energy from the system, reversing the growth and causing $\delta g$ to shrink.  Once $\delta g$ is sufficiently small, the drive once again causes linear growth.

 Of course, the limit $\alpha = 0$ is somewhat academic: Fig.~\ref{drivenFigres}(a) shows the generic behavior that we find when $\alpha\neq 0$.
We see only a single beat, whose physical origin is very different from the $\alpha=0$ case.  Here what happens is that as $\delta g$ grows we develop strong occupation of the non-condensed modes.  The resulting Bose enhancement means that the decay rate exceeds the drive-induced growth rate, causing $\delta g$ to fall. {The time scale where this effect becomes important is set by the inverse driving rate $1/\delta \epsilon$}.   At long times one finds small oscillations which are largely due to coherent oscillations between condensed and non-condensed modes.

 Figure~\ref{drivenFigres}(b) shows the occupation of the non-condensed modes during this process.  The population is mainly excited in a narrow range of $\epsilon_k$, corresponding to $\Omega_k\sim \omega_g$ (cf. the  discussion following Eq.~(\ref{OmegakFk})).  Figure ~\ref{drivenFigres}(c) shows the population at this resonant energy.  There is a remarkable enhancement compared with the undriven situation.  The oscillations again indicate coherent interconversion between condensed molecules and non-condensed atom pairs.

\begin{comment} even for a small damping $\alpha \ll \delta_\epsilon$, we observe a strong suppression of the oscillations $\delta g$ after the initial evolution, and no recurrences, see Fig.\ \ref{drivenFigres}(a). The most important reason for this suppression is not the anharmonicity of the effective potential, but rather the accelerated decay of the resonant continuum mode with $\Omega_k = \omega_g$: the external drive initially enhances the fluctuations $\delta g$. These oscillations strongly drive the resonant mode via Eq.\ \eqref{deltaveq}, as visualized in Figs.\ \ref{drivenFigres}(b),(c). In turn, the enhanced mode population leads to a Bose-enhanced accelerated decay of the $\delta g$ oscillations, escalating the analogous effect observed in the equilibrium case. The initial growth of the $\delta g$ amplitude is therefore rapidly cut off, which is in qualitative agreement with measured trajectories in the experiment \cite{chicagoexp}. At long times, only small forced oscillations without resonant enhancement persist. 
\end{comment}

%
%While the continuum coupling strongly suppresses the coherent oscillations in the studied driving scheme, this is not necessarily the case for other driving mechanisms. In particular, a periodic modulation of the coupling $\alpha \rightarrow \alpha \cos(\omega_\alpha t)$ itself can lead to a resonant oscillation enhancement when $\omega_\alpha = \omega_g$, as numerically demonstrated in App.\ \ref{stochasticApp}. This is likely tied to the phenomenon of stochastic resonance: for two-level systems, it is known that amplitude noise can enhance the effect of a periodic modulation in the tunneling rate \cite{Wellens2004, PhysRevLett.101.200402}.  However, in the atom-molecule conversion problem implementation of such a coupling-constant drive requires an involved experimental beyond Feshbach resonance tuning, e.g. a modulation of the trapping potential. 

\section{Experimental Considerations}\label{sec:exp}

\subsection{Energy and Timescales}
In this paper we work in natural units, which make the phenomena most transparent.  To connect with experiments, we now re-introduce physical units and provide  estimates for typical parameter values.

Solving the 2-body problem with the Hamiltonian in Eq.~\eqref{basichamil} gives a scattering length \cite{timmermans}
$a_s=-{m \lambda^2}/({4\pi \hbar^2 \epsilon})$,
which should be matched with the phenomenological behavior of $a_s$ close to a Feshbach resonance, $a_s \simeq -a_{\rm bg}\Delta B/(B-B_0)$.  %{Our model only applies near resonance, so the first term can be neglected, and we use $a_s=- a_{\rm bg} \Delta B/(B-B_0)$.}

The experiments work with $^{133}$Cs, for which $m=133$amu.  According to \cite{chicagoexp,wang}, the Feshbach resonance is at $B_0=19.849(2)$G, and has a width $\Delta B=8.3(5)$mG.  Far from the resonance the low-energy scattering is characterized by a background scattering length $a_{\rm bg}=163 a_\text{B}$, where $a_\text{B}$ is the Bohr radius.
The magnetic moment difference between the closed (molecular) and open (atomic) channel is  
$\delta\mu=\hbar\times4.8(2)(\mu $s$)^{-1}$%MHz\,
G$^{-1}$ The molecular energy is then $\epsilon=\delta \mu (B-B_0)$.  Thus we conclude that 
$\lambda=\sqrt{(4\pi \hbar^2 \delta \mu a_\text{bg}\Delta B)/m }$. A typical density reported in the experiments is $n = 2.9\times10^{13}$cm$^{-3}$. Inserting these values into the definition of the effective dimensionless coupling $\alpha$, Eq.\ \eqref{pncons}, the  unit of time $\hbar/(\sqrt{2n} \lambda)$ and magnetic-field equivalent of the unit of energy, $\sqrt{2n} \lambda/\delta \mu$, leads to parameter values as shown in Table \ref{table1}.

\begin{table}[H]
\centering
\begin{tabular}{|c |c |c|}
\hline
 \ $\boldsymbol{\alpha}$ \ & \ \textbf{time unit} \  & \ \textbf{(energy unit)}/$\boldsymbol{\delta \mu}$ \   \\ % inserts table %heading
\hline
0.27& $\tau=$0.09 ms& $\delta B/\epsilon=$2.3 mG \\
 [1ex]
\hline
\end{tabular}
\caption{Typical parameter values for comparison with experiments\ \cite{chicagoexp,wang}.   The first column shows the small parameter in our expansion, $\alpha$, which characterizes the coupling between atoms and molecules.  The second shows $\tau$, the time unit used in all graphs.  The third gives the conversion factor which relates the dimensionless molecular energy $\epsilon$ to the physical magnetic field detuning from resonance, $\delta B = B - B_0$.}
\label{table1}
\end{table}

\subsection{Experimental Protocols}
\label{sec:protocol}
The experimental protocol used in Refs.~\cite{chicagoexp,wang} throws the system far out of equilibrium.  This causes large signals, but makes modeling challenging.  Small perturbations can be amplified through the non-equilibrium dynamics and there are no small parameters controlling the behavior.  We argue that the core physics can be explored using more robust near-equilibrium experiments.  For detailed modeling of the current experiments we refer the reader to Ref. \cite{wang}.

We advocate for a quasi-equilibrium experiment, where one slowly ramps the magnetic field to produce a setting where the atoms and molecules are in chemical equilibrium.  One could then perturb the magnetic field to excite the oscillations studied in Sec.~\ref{linac} or \ref{drivingsec}, depending on if the perturbation is static or periodic. 

One additional feature is that, as noted in  \cite{wang,basu,timmermans}, near the resonance the equilibrium system's compressibility is negative unless the molecule-molecule scattering is  large.  Consequently there is a mechanical instability which will lead to collapse \cite{Donley2001,hulet}.  It would be extremely interesting to study this instability in the strongly interacting limit, and the interplay between chemical and mechanical dynamics.  If one wants to avoid this complication, one can restrict experiments to  detunings where the system is mechanically stable or where the collapse time is long.

\section{ Summary and Outlook}
\label{conclusionsec}

 Some of the most important questions being addressed in atomic, molecular, and optical systems involve the coupling between macroscopic quantum degrees of freedom and incoherent baths.  These questions are key to manipulating quantum information \cite{harrington}, and exploring the ways in which quantum systems thermalize \cite{Ueda2020}.  In this work we explored a dramatic example of such dynamics, namely the interconversion between atomic an molecular Bose-Einstein condensates near a Feshbach resonance.  We have shown that the coherent atom-molecule oscillations are damped by the decay of molecules into non-condensed atomic pairs.  This is a resonant process, and the atomic pair energy is set by the atom-molecule oscillation frequency.  This pair production is somewhat analogous to the ``atomic fireworks'' in driven condensates \cite{BoseFire}.  Here the atom-molecule oscillations play the role of the drive.

A key feature of this system is that the decay process is non-linear.  Due to Bose stimulation, it occurs faster when there are more non-condensed pairs in the system.  This leads to significant deviation from simpler theories which assume a Markovian, memory-free, bath.

Motivated by the experiments \cite{chicagoexp}, we also considered the response of this system to periodically modulating the molecular binding energy.  Such modulation can drive atom-molecule oscillations, but coupling to the continuum causes these oscillations to damp out. Again, the non-linear nature of the decay process is significant, and they have a larger impact for resonant driving.  

 Our work complements the recent calculations of Wang et al., who used a similar formalism to study the far-from equilibrium dynamics after a magnetic field quench \cite{wang},  mimicking the experimental protocol.  They make direct comparison with the experiment, revealing both the strengths and weaknesses of this type of Bogoliubov description.  We argue that better quantitative agreement will be found in near-equilibrium experiments, and advocate for a modified experimental approach.

  We made several approximations in our analysis.  Most crucially, we included processes in which condensed molecules could dissociate into atomic pairs, but did not include terms where those atoms scattered off  one-another, nor where they recombined to form molecules with non-zero momentum.  At longer times one would expect these neglected terms to become important.  For example, they are required for the eventual thermalization of the system.  The first step towards modeling this longer-time physics involves extending our ansatz to include molecule-molecule and atom-molecule pairs \cite{threebody}.
Additionally, it would be natural to model the 3-body processes which appear to be relevant in the experiment \cite{chicagoexp}.

 Finally it would be interesting to analyze whether the noise generated by the interaction with the continuum  is always detrimental to an external drive of the condensate, or whether driving protocols exist where the noise can be beneficial in the sense of a stochastic resonance \cite{PhysRevLett.101.200402}. 
\\ \textit{Acknowledgements}: 
We thank Chin Cheng for clarifications about the experiment, and Tin-Lun Ho for discussions, including suggesting the ansatz which we used. 
D.P.\ acknowledges funding by the German Research Foundation (DFG) under Project-ID 442134789, and E.J.M. acknowledges support from the National Science Foundation under Grant No.\ PHY-2110250.

\appendix 
\section{Pair coherent states}
\label{cohapp}

In this Appendix, we derive the relevant properties of the coherent state $\ket{\Psi}$, Eq.\ \eqref{newpsi}. First, we recap the properties of the condensate parts, which fulfill 
\begin{align} 
\braket{\Psi| \psi_0 | \Psi} &= \sqrt{N} f e^{i\theta} \\ \notag
\braket{\Psi| \psi_0^\dagger | \Psi} &= \sqrt{N} f e^{-i\theta} \\ \notag
\braket{\Psi| \psi_0^\dagger \psi_0 |\Psi} &= N f^2  \\ \notag
\braket{\Psi| \psi_0 \psi_0 |\Psi} &= N f^2 e^{2i\theta} \ . 
\end{align} 
We now derive  analogous properties for the single-mode state \cite{nozieres1982particle}
\begin{align} 
\ket{\tilde \Psi_k} = \exp(z_k \psi^\dagger_k \psi^\dagger_{-k} ) \ket{\text{vac}} = \sum_{n = 0}^\infty z_k^n \ket{n,n} , 
\end{align} 
where $\ket{n,n}$ is a number eigenstate with $n$-fold populated atomic states with momenta $k$ (first entry) and $-k$ (second entry). The squared norm of $\ket{\tilde \Psi_k}$  is found by summing a geometric series,
\begin{align}
\braket{\tilde \Psi_k| \tilde \Psi_k} = \frac{1}{1 - |z_k|^2} \ . 
\end{align} 
We then note that 
\begin{align} 
\psi^\dagger_k \psi^\dagger_{-k} \ket{ \tilde \Psi_k} = \sum_{n = 0}^\infty (n+1) z_k^n \ket{n+1,n+1} = \partial_{z_k} \ket { \tilde \Psi_k} , 
\end{align}
which implies 
\begin{align} 
\braket{\tilde \Psi_k | \psi^\dagger_k \psi^\dagger_{-k} |\tilde\Psi_k} =
\partial_{z_k}\braket{\tilde \Psi_k| \tilde \Psi_k}=
\frac{\overline{z}_k}{(1 - |z_k|^2)^2}  \ , 
\end{align} 
where the derivative is taken holding $\overline{z}_k$, the complex conjugate of $z_k$, fixed.
As a result, the normalized state $\ket{\Psi} =\braket{\tilde \Psi_k| \tilde \Psi_k}^{-\frac{1}{2}} \ket{\tilde \Psi}$ fulfills 
\begin{align} 
\braket{\Psi_k | \psi^\dagger_{k} \psi^\dagger_{-k} |\Psi} = \frac{\overline{z}_k}{(1 - |z_k|^2)} \ . 
\end{align} 
The first identity in Eq.\ \eqref{Psiidentities} follows by pulling out a phase factor, and the second one can be derived in the same vein.

\section{Derivation of the time evolution}
\label{diffeqapp}

To obtain the main differential equation, Eq.~\eqref{fulltwo}, from the coherent state representation, one needs to  calculate the Lagrangian $\mathcal{L} = \braket{\Psi(t) | i \partial_t - H | \Psi(t)}$ and find its stationary point. The expectation value $\braket{\Psi(t) | H | \Psi(t)}$ is given by Eq.\ \eqref{Ewithalpharen} with time-dependent coefficients. To find the part $\braket{\Psi(t) | \partial_t | \Psi(t)}$, we first consider $\ket{z_f} \equiv \exp\left(- N f^2/2\right) \times \notag \exp\left(z_f\sqrt{N} \psi^\dagger_0 \right) \ket{\text{vac}}$, which is one of the factors entering $\ket{\Psi(t)}$. Recall that $z_f = f e^{i\theta}$. We have 
\begin{align}
\label{tdcoh1}
\partial_t \ket{z_f} &= \left(\dot{z}_f \partial_{z_f} + \dot{\bar{z}}_f \partial_{\bar{z}_f} \right) \ket{z_f} 
\end{align} 
and
\begin{align} 
\partial_{\bar{z}_f} \ket{z_f} &= - \frac{N z_f}{2} \ket{z_f}  \\ \notag
\partial_{z_f} \ket{z_f} &= \left(- \frac{N \bar{z}_f}{2} + \sqrt{N} \psi^\dagger_0 \right) \ket{z_f}  \\ \notag
\braket{z_f | \partial_{\bar{z}_f} | z_f} &= - \frac{N z_f}{2}  \\ \notag
\braket{z_f | \partial_{{z}_f} | z_f} &=  \frac{N \bar{z}_f}{2}  \ . 
\end{align} 
As a result: 
\begin{align} 
\braket{z_f| \partial_t | z_f} &= \frac{N \bar{z_f}}{2} \dot{z}_f  - \frac{N {z_f}}{2}  \dot{\bar{z}}_f  \to  N \bar{z}_f \dot{z}_f  , 
\end{align} 
where we have integrated by parts (with respect to the integration $S = \int dt \mathcal{L}$). The contributions of the remaining parameters $z_g, z_k$ are obtained in the same manner. Eq.\ \eqref{fulltwo} can then be obtained from the conditions $\partial_{\bar{z}_i} \mathcal{L} = 0, i = \{f,g,k\}$ and measuring the time in units of $1/(\sqrt{2n} \lambda)$.

\section{Properties of soliton-like solutions} 
\label{solitonapp} 
The soliton-like solutions to Eq.\ \eqref{fulltwo} obey the boundary conditions $f(t = 0) = 1, g(t = 0) = 0^+$. When the continuum modes are neglected ($\alpha = 0$),  they evolve along a contour of vanishing condensate energy $E_\text{cond} = 0$ [Eq.\ \eqref{Energydensity}]. This condition can be used to eliminate the phase variable $\eta$ and obtain a single differential equation for $g$: 
\begin{align} 
\dot{g} = \sqrt{(1 -g^2)^2 - (g\epsilon/2)^2} \ . 
\end{align} 
From $\dot{g} = 0$ we obtain the turning point value $g_\text{max} = \sqrt{ 1+ (\epsilon/4)^2} - \epsilon/4$. Separating variables, the oscillation period can be extracted as 
\begin{align} 
T = 2\int_0^{x_\text{max}} dx \frac{1}{2\sqrt{(1-x^2)x - \left(\frac{x \epsilon}{4}\right)^2}}  \ ,  
\end{align} 
where $x = g^2$, $x_\text{max} = g_{\text{max}}^2$. Integration gives 
\begin{align}
T = \frac{\text{EllipticK}\left(\frac{1}{1 + (\epsilon/4)^2}\right)}{\sqrt{1 + (\epsilon/4)^2}} \sim \begin{cases} \log(1/\epsilon ) {\quad \epsilon\to 0}\\ 1/\epsilon {\qquad\quad\epsilon\to\infty}
\end{cases}, 
\end{align} 
where EllipticK is the complete elliptic integral of the first kind.

\section{Mean field vs.\ exact dynamics}
\label{beyondmfapp}

In this Appendix, we compare the dynamics obtained from the mean-field-like coherent state variational approach with exact diagonalization results for the pure condensate case ($\alpha = 0$). For fixed total particle number $N$, the  exact wavefunction can be expressed as
\begin{align} 
\label{finiteNspan}
\ket{\psi} = \sum_{m = 1}^{N/2} c_m(t) \ket{N_\phi=m,N_\psi=N-2m}, 
\end{align}
where $\ket{N_\phi=m,N_\psi=N-2m}$ is a number eigenstate that contains $m$ molecules and $N - 2m$ atoms. For initial conditions we choose coefficients $c_m$ such that the system is in a finite-$N$ version of the coherent ansatz \eqref{coh1}, with $\braket{\psi  | \phi^\dagger_0 \phi_0 | \psi }^{-1/2} \equiv g(0) \sqrt{N/2}$. We then compare the evolution of $g_N(t) \equiv \sqrt{ \frac{2}{N}} \braket{ \psi  | \phi^\dagger_0 \phi_0 | \psi }^{-1/2}$ with $g(t)$ obtained from the $(N = \infty)$ coherent state. At short times, the coefficients $c_m$ are sharply peaked around the mean-field result. At longer times, dephasing sets in: the $c_m$-distribution broadens since different quantum trajectories effectively evolve with different frequencies. This causes $g_N(t)$  to deviate from $g(t)$.

\begin{figure}
\centering
\includegraphics[width=\columnwidth]{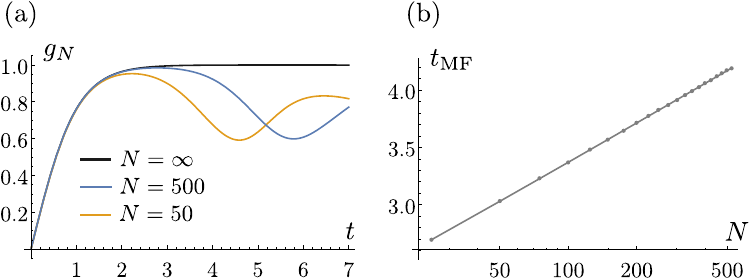}
\caption{Exact vs mean-field dynamics for soliton-like solution with $g(0) = 0^+$. (a) $g(t)$ for fixed total particle number $N$ (mean-field corresponds to $N = \infty$). (b) Scaling of $t_{\text{MF}}$ with $N$ on a log-linear plot.}
\label{exactsol}
\end{figure}

In Fig.\ \ref{exactsol}, we show results for the soliton-type trajectory with $g(0) = 0^+$. While the finite-$N$ trajectories follow the mean-field result accurately for $g \ll 1$,  they deviate significantly as $g\to 1$ [Fig.\ \ref{exactsol}(a)]. We  define $t_{\text{MF}}$ as the time when the mean field and the finite-$N$ trajectories deviate by $10\%$ (a value which is chosen arbitrarily). As seen in the log-linear plot shown in Fig.\ {\ref{exactsol}}(b), $t_{\text{MF}}$  grows as $\log(N)$ for the soliton trajectory \cite{PhysRevLett.86.568,PhysRevA.64.063611}.

\begin{figure}
\centering
\includegraphics[width=\columnwidth]{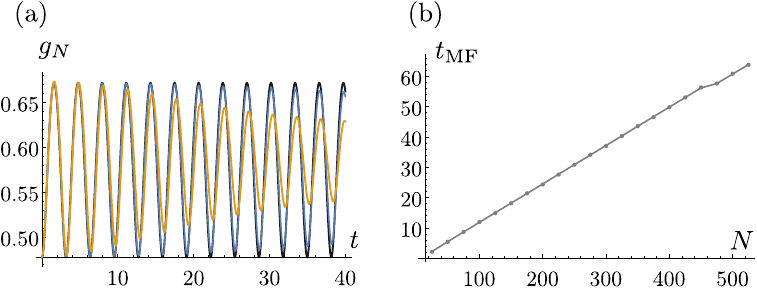}
\caption{Exact vs mean-field dynamics for oscillatory solution with $g = g^\star - 0.04$. (a) $g(t)$ for fixed total particle number $N$, with colors as in Fig.\ \ref{exactsol}.  (b) Scaling of $t_{\text{MF}}$ with $N$. }
\label{exactosc}
\end{figure}

In Fig.\ \ref{exactosc}, we show analogous results for an oscillatory trajectory with $g(0) \lesssim g^\star$. At fairly small $N$, dephasing leads to damped oscillations, see Fig.\  \ref{exactosc}(a). However, at larger $N$ the mean-field result quickly becomes accurate, and the corresponding dephasing time $t_\text{MF}$ scales linear in $N$. For ultracold experiments with $O(10^4)$ atoms \cite{chicagoexp}, the mean field approximation is excellent  for all experimentally relevant times.

\section{Evaluation of decay rate}
\label{rateapp}

According to Eq.\ \eqref{Gammaintro}, the relevant integral for the determination of the damping rate is of the form  
\begin{align} 
\label{Iapp}
I \equiv  \int_k \dot{\delta v_k} , 
\end{align} 
plus a  contribution from $u_k$ which can be evaluated in the same manner. Consider the solution for $\delta v_k$ close to resonance,  $\Omega_k \simeq \omega_g$ (we choose our origin of time so that $\delta g$ contains no $\sin(\omega_gt)$ terms): 
\begin{align} 
&\delta g(t)\bigg|_{\alpha = 0} = A \cos(\omega_g t) \\  
\label{vresapp}
&\delta v_k^\text{res} (t) =   \frac{A F_k}{2(\Omega_k - \omega_g)} \left( \sin(\omega_g t) - \sin(\Omega_k t) \right),  
\end{align} 
where $F_k, \Omega_k$ are given in Eq.\ \eqref{OmegakFk}. For $t \gg 1/\omega_g$,  Eq.\ \eqref{vresapp} is strongly peaked at $\Omega_k = \omega_g$, with a peak height that is linear in $t$. 
In Eq.\ \eqref{Iapp}, it is convenient to change variables as $\epsilon_k  \rightarrow \Omega_k$, yielding a modified integration measure $h(\Omega_k)$: 
\begin{align} 
d\epsilon_k \sqrt{\epsilon_k} \rightarrow d\Omega_k \frac{\Omega_k \sqrt{\sqrt{4g^2 + \Omega_k^2} - 2g}}{2\sqrt{2} \sqrt{ 4g^2 + \Omega_k^2}}  \equiv d\Omega_k h(\Omega_k)  \ . 
\end{align} 
Since we are interested in the resonant contribution only, we restrict the integral to a symmetric region around $\Omega_k = \omega_g$: 
\begin{align} 
I_\text{res} \equiv \int_{\omega_g - \kappa}^{\omega_g + \kappa}  d\Omega_k f(\Omega_k) \frac{\sin(\omega_g t) - \sin(\Omega_k t)}{\Omega_k - \omega_g} , 
\end{align} 
where $\kappa \ll \omega_g$, and $f(\Omega_k) = A F_k h(\Omega_k)/2$ contains all terms that are non-singular when $\Omega_k \rightarrow \omega_g$. Changing variables to $x = (\Omega_k - \omega_g)t$, we can rewrite $I_\text{res}$ as  
\begin{align} 
\notag
I_\text{res} &= \text{Re} \left[ e^{i\omega_g t}  \int_{-\kappa t}^{\kappa t} dx \frac{1-e^{ix}}{x} f\left(\frac{x}{t}+ \omega_g \right) \right] \\ \notag
&\simeq \text{Re} \left[ e^{i\omega_g t}  \int_{-\kappa t}^{\kappa t} dx \frac{1-e^{ix}}{x} f\left( \omega_g \right) \right] \\ \notag &= \text{Re} \left[-i e^{i\omega_g t}  \int_{-\kappa t}^{\kappa t} dx \frac{\sin(x)}{x} f\left( \omega_g \right) \right]   \\ \notag  &\overset{\kappa t \gg 1}{\simeq} \pi f(\omega_g) \sin(\omega_g t)  \\ &= 
- \frac{\pi}{2 \omega_g} F_k h(\Omega_k)\bigg|_{\Omega_k = \omega_g} \dot{\delta g}(t) \bigg|_{\alpha = 0} . 
\end{align} 
The prefactor in this result determines the damping rate.

{\section{Three-body interactions}
\label{app:3body}

To include three-body processes that involve the decay of a molecule and an atom into three atoms and vice versa, the Hamiltonian can be extended with a term \cite{chicagoexp}

\begin{align}
\label{Hint3}
&H_\text{int}^{(3)} =  
\\ &\frac{\lambda_3}{V^{3/2}} \sum_{kpqs} \phi_{k+p+q-s}^\dagger \psi_s^\dagger \psi_k \psi_p\psi_q
+ \psi_k^\dagger \psi_p^\dagger\psi_q^\dagger \psi_s\phi_{k+p+q-s}, \notag
\end{align} 
with an intensive short-range coupling constant $\lambda_3$. 

We focus on a simplified situation where the three-body term dominates over the two-body one, $\lambda_3 \gg \lambda$, and measure energies in units of $\sqrt{2} n^{3/2} \lambda_3$. First, we consider the condensate-only case, where we reproduce the two-body results qualitatively, with only a difference in numbers. The energy takes the form 
\begin{align} 
E^{(3)}_\text{cond} = \frac{\epsilon}{2} g^2 + \cos(\eta) g f^4 = \frac{\epsilon}{2} g^2 + \cos(\eta) g (1 - g^2)^2 \ . 
\end{align} 
At vanishing detuning, the energy minimum corresponds to  $\eta^\star = \pi, g^\star = 1/\sqrt{5}$. The equations of motion read 
\begin{align}
\partial_t g &= (1-g^2)^{2} \sin\eta\\ \notag
\partial_t \eta &= \epsilon +
\left(\frac{(1-g^2)^{2}}{g}-4 g(1-g^2)\right)\cos\eta\ .
\end{align}

Similar to the two-body case, these equations admit soliton-like solutions for $g(0) \simeq 0$ and oscillatory ones for $g(0) \simeq g^\star$. The scaling of energies (e.g, the oscillation frequency) with density as $n^{3/2}$, possibly renormalized by trap effects,  and the value of $g^\star = 1/\sqrt{5}$  suggest that the recent experiment \cite{chicagoexp} was dominated by three-body processes.  

To study the impact of the atomic continuum, we can evaluate the three-body Hamiltonian using the general ansatz $\ket{\Psi}$ from Eq.\ \eqref{newpsi}. We obtain an expectation value 
\begin{align} \notag
&E^{(3)} = \braket{\Psi|H|\Psi}/ (\sqrt{2} n^{3/2} \lambda_3 N) =  
 \frac{\epsilon}{2} g^2 + \cos(\eta) f^4 g + \\ &+  \alpha_3 \int_k \frac{1}{1 - |z_k|^2} \left[3 f^2 g \left( u_k + \cos(\eta) |z_k|^2\right) + \epsilon_k  |z_k|^2  \right] , 
 \label{3benergy}
\end{align}
 where $\alpha_3 = m^{3/2} \left( \lambda_3 \sqrt{2} n^{3/2} \right)^{3/2}/(\sqrt{2} n \pi^2)$, compare Eq.\ \eqref{alphadef}.
It is clear that the minumum occurs when $\eta = \pi$, $z_k = u_k < 0$.
  The factor of $3$ in the second line of Eq.\ \eqref{3benergy} is crucial: It represents the different possibilities to choose two non-condensed atoms out of the interaction term \eqref{Hint3} in the pairing or density channel.

Minimizing Eq.~(\ref{3benergy}) is harder than in the case of two-body interactions in Eq.~(\ref{Ewithalpha}), as the interconversion term contains the factor $f^2 g$.  Thus using number conservation to eliminate $f$ results in quite involved expressions.  One can proceed by introducing a Lagrange multiplier, but it 
 is more elegant to further restrict our ansatz, and produce an upper bound to the ground state energy.  To simplify the algebra we consider the resonant case, $\epsilon=0$.  
 
For our restricted ansatz we take
 $u_k=0$ when $\epsilon_k$ is above some cutoff $\xi$.  We choose $\xi$ to be sufficiently small that we can neglect the $\epsilon_k$ term in Eq.~(\ref{3benergy}).  Because our approach is variational, this gives us an upper bound to the energy, and making $\xi$ smaller should increase our energy.

We then use that
$|u_k| > u_k^2$, to bound the energy as

\begin{align} 
\label{E3bound}
E^{(3)}_\text{min} \leq  - g(f^4  + 6 f^2   c^2)+{\cal O}(\xi)\ , 
\end{align} 
 where 
 \begin{align} 
 c^2 \equiv \alpha_3 \int_k \frac{u_k^2}{1 - u_k^2} 
 \end{align} 
 is the total number of continuum states, and we have taken $\eta=\pi$. For the chosen configuration with $u_k = z_k$, number conservation implies 
 $g^2 + f^2 + c^2 = 1$. Inserting this into  \eqref{E3bound}, we find $(g^\star)^2 = 1/5, (c^\star)^2 = 8/25, (f^\star)^2 = 12/25$, which leads to an energy bound $E_\text{min}  \leq - 0.515$, significantly lower than the estimate with $c^2 = 0$ for which  $E = -0.268$. 

 We emphasize that even for an infinitesimal $\alpha_3$, this ansatz creates an order 1 change in the energy, and leads to a  
%As a result, even in the limit $\alpha_3 \searrow 0$, the energy minimum contains a 
macroscopic change in the number of condensed particles:  $(f^*)^2$ is reduced from the value of $4/5$, which is found if one takes $\alpha_3=0$.
These particles can be thought of as `pairs' -- meaning that they have correlated momenta, but are not molecules.  Note that these non-condensed pairs all have $\epsilon_k<\xi$, for infinitesmal $\xi$.  Thus this represents a form of pair condensation or a fragmented condensate \cite{nozieres1982particle,PhysRevA.74.033612}.

Pair condensates are quite exotic, and in the absence of an atomic condensate would have vortex structures with an atomic phase winding of $\pi$, corresponding to a half-quantum vortex \cite{PhysRevLett.92.160402,PhysRevLett.93.020405}.  Models with pair condensation often have mechanical instabilities which prevent their observation \cite{nozieres1982particle,ejmcollapse}, and more analysis is needed to see if this physics is observable here.

To verify that this analysis is sensible, we
apply the same procedure to the two-body case in Eq.~(\ref{Ewithalpha}).
We find an upper bound 
\begin{align} 
E^{(2)}_\text{min} \leq - g(f^2 + c^2),
\end{align} 
and it is not energetically favorable to convert atoms into non-condensed pairs.

}

{\section{Coherent state description of excited molecular states}
\label{excitedmolapp}

To include the possible decay into excited molecular states in the two-body case, the coherent state ansatz of Eq.\ \eqref{newpsi} needs to be generalized. One possible  ansatz (not normalized) is
\begin{align} 
\notag
&\ket{\tilde\Phi} = \prod_{k}{\vphantom{\sum}}'  \ket{\tilde\Phi_k} ,  \\ 
\label{Phikansatz}
&\ket{\tilde\Phi_k} = \exp\left( \eta_k ( \phi^\dagger_k \psi^\dagger_{-k} + \phi^\dagger_{-k} \psi^\dagger_k) + z_k \psi^\dagger_{k} \psi^\dagger_{-k} \right) \ket{\text{vac}} \ ,
\end{align} 
where have suppressed the condensate operators, and have not pulled out any phase factors.
%at this stage to shorten the derivations. 
Note that the most general coherent ansatz should also contain a term $\phi^\dagger_{k} \phi^\dagger_{-k}$. For simplicity, we restrict ourselves to the limited ansatz \eqref{Phikansatz} instead, since it captures the essential physics and should be quantitively accurate for small molecule populations, as it contains the relevant linear terms in $\phi_k^\dagger$.

The wavefunction,
$\ket{\tilde\Phi_k}$, can be expressed as 
\begin{align} 
&\ket{\tilde\Phi}_k =  \\  \notag &\sum_{l,m,n = 0}^\infty \eta_k^{l+m} z_k^n \sqrt{\binom{m+n}{m}\binom{l+n}{l}} \ket{ l, m, m+n, l + n} \ , 
\end{align} 
where 
\begin{align}
&|M_k,M_{-k},N_k,N_{-k}\rangle = \\ & \notag
 \frac{(\phi_k^\dagger)^{M_k}
(\phi_{-k}^\dagger)^{M_{-k}}(\psi_k^\dagger)^{N_k}
(\psi_{-k}^\dagger)^{N_{-k}}}{\sqrt{M_k! M_{-k}! N_k! N_{-k}!}}|{\rm vac}\rangle.
\end{align}
I.e., the first two entries refer to the population of molecules with momenta $k, -k$ and the last two to atoms, again with $k, -k$. %(schematically: $\ket{\phi_k, \phi_{-k}, \psi_{k}, \psi_{-k}}$). 
The squared norm of $\ket{\Phi_k}$ is given by 
\begin{align} 
\braket{\tilde\Phi_k|\tilde\Phi_k} = \sum_{l,m,n= 0}^\infty |\eta_k|^{2(l+m)} |z_k|^{2n} \binom{m+n}{m} \binom{l+n}{l} \ . 
\end{align} 
This sum can be evaluated with help of the identities 
\begin{align} 
\label{binomialidentities}
&\binom{m+n}{m} = (-1)^m \binom{-(n+1)}{m}\  \notag , \\
&\quad \sum_{m = 0}^\infty \binom{a}{m} x^m = (1+x)^a \ . 
\end{align}  
We obtain the compact expression
\begin{align} 
\braket{\tilde\Phi_k|\tilde\Phi_k} = \frac{1}{(1-|\eta_k|^2)^2 - |z_k|^2} 
\end{align} 
We can now introduce the normalized state $\ket{\Phi_k} = \braket{\tilde\Phi_k|\tilde\Phi_k} ^{-1/2} \ket{\tilde\Phi_k}$, and evaluate the expectation values of various operator combinations appearing in the Hamiltonian. Employing identities similar to 
\eqref{binomialidentities}, we obtain 
\begin{align} 
\braket{\Phi_k | \phi^\dagger_k \phi_k |  \Phi_k} &= \frac{|\eta_k|^2(1-|\eta_k|^2)}{(1-|\eta_k|^2)^2 - |z_k|^2}  \\  \notag
\braket{\Phi_k | \psi^\dagger_k \psi_k |  \Phi_k} &= \frac{|\eta_k|^2(1-|\eta_k|^2) + |z_k|^2}{(1-|\eta_k|^2)^2 - |z_k|^2} \\ \notag
\braket{\Phi_k | \psi_k \psi_{-k} |  \Phi_k} &= \frac{z_k}{(1-|\eta_k|^2)^2 - |z_k|^2} \\ \notag
\braket{\Phi_k | \phi^\dagger_{k} \psi_k |  \Phi_k} &= \frac{\overline{w}_k z_k} {(1-|\eta_k|^2)^2 - |z_k|^2}  \ . \\ \notag
\end{align} 
Relabeling the parameters as $z_k \rightarrow \exp(i\chi) z_k,\   \eta_k \rightarrow \exp(i\gamma) \eta_k$ for convenience (with $\gamma$ to be determined) and reintroducing the condensate fractions, the normalized energy expectation value $E =   \braket{\Phi | H | \Phi}/(\sqrt{2 n}  \lambda N)$ can be written as 
\begin{align} 
\label{bestenergy}
E &= \frac{\epsilon}{2} g^2  + \cos(\eta) g f^2\ \\
\notag
& +  \alpha \int_k \epsilon_k \frac{ \frac{3}{2} |\eta_k|^2 (1-|\eta_k|^2) + |z_k|^2}{ (1 - |\eta_k|^2)^2 - |z_k|^2}     \\ 
& +  \alpha \int_k  \frac{g u_k + \sqrt{2} f \text{Re}\left[\exp(i(\theta + \chi - \gamma )) z_k \overline{\eta}_k\right]}{ (1 - |\eta_k|^2)^2 - |z_k|^2}  \notag \     \\  \label{molappcons}
f^2 &= \left( 1 - g^2 - \alpha \int_k \frac{ 3 |\eta_k|^2 (1-|\eta_k|^2) + |z_k|^2}{ (1 - |\eta_k|^2)^2 - |z_k|^2)} \right) . 
\end{align} 
where Eq.\ \eqref{molappcons} shows the particle conservation. 

Using a similar argument to 
App.\ \ref{app:3body}, we can again show that the minimum contains a pair condensate.  As discussed in that section, it is not clear if that state is robust against mechanical instabilities, without adding extra terms to the Hamiltonian.

}

\bibliography{molecule}
\bibliographystyle{apsrev4-2}

\end{document}